# Spectroscopic study of impurities and associated defects in nanodiamonds from Efremovka (CV3) and Orgueil (CI) meteorites

A. A. Shiryaev[1,*], A. V. Fisenko[2], I. I. Vlasov[3], L. F. Semjonova[2], P. Nagel[4], S. Schuppler[4]

1) Institute of Physical Chemistry and Electrochemistry RAS, Leninsky pr. 31, Moscow 119991, Russia
2) Vernadsky Institute of Geochemistry and Analytical Chemistry RAS, Kosygina Str. 19, Moscow, Russia
3) General Physics Institute RAS, Vavilov Str., 38, 119991 Moscow, Russia
4) Karlsruhe Institute of Technology, Institute für Festkörperphysik, Karlsruhe 76133, Germany

* Corresponding author: shiryaev@phyche.ac.ru

**Abstract –** The results of spectroscopic and structural studies of phase composition and of defects in nanodiamonds from Efremovka (CV3) and Orgueil (CI) chondrites indicate that nitrogen atomic environment in meteoritic nanodiamonds (MND) is similar to that observed in synthetic counterparts produced by detonation and by the Chemical Vapour Deposition (CVD)-process. Most of the nitrogen in MND appears to be confined to lattice imperfections, such as crystallite/twin boundaries and other extended defects, while the concentration of nitrogen in the MND lattice is low. It is suggested that the N-rich sub-population of MND grains may have been formed with high growth rates in environments rich in accessible N (i.e., N in atomic form or as weakly bonded compounds). For the first time the silicon-vacancy complex (the "silicon" defect) is observed in MND by photoluminescence spectroscopy.

## 1. INTRODUCTION

Nanodiamonds in primitive carbonaceous chondrites contain high nitrogen contents (1–4 wt%) (Russell et al., 1996; Newton et al., 1995). The main peak of N release from samples of meteoritic nanodiamonds (MND) during step oxidation coincides with that of carbon (450–550 ˚C), thus strongly favoring the hypothesis that the N resides in diamond grains (Lewis et al., 1987; Newton et al., 1995; Russell et al., 1996; Verchovsky et al., 1998). Russell et al. (1996) reported a correlation between the nanodiamond content in meteorites and the N concentration in the MND. Low C/N ratio (25–100) in diamonds from the least metamorphosed chondrites (Acfer 094, ungrouped Orgueil, CI) suggests the existence of N-enriched (up to several wt% N) population of MND (Russell et al., 1996; Fisenko and Semjonova, 2006). It has been suggested (Fisenko et al., 1992) that thermal metamorphism of the meteorite body results in preferential destruction of the N-rich nanodiamonds and increase of C/N ratio in diamonds (e.g., up to 242±32 in Indarch (EH4), Russell et al., 1996). This scenario is supported by the higher oxidation rate of macrodiamonds with nitrogen concentrations exceeding a threshold value of ~1000 ppm (Zhdankina et al., 1986).

It has been demonstrated experimentally that the efficiency of N incorporation into a diamond lattice during the Chemical Vapour Deposited (CVD-process) usually does not exceed $10^{-3}$, thus making impossible significant levels of diamond doped by nitrogen (Jin and Moustakas, 1994; Samlenski et al., 1995). The concentration of nitrogen in diamond grains in the CVD films rarely exceeds 200–300 ppm and is largely determined by the stability of nitrogenous compounds present in the growth chamber.

A much higher N content (up to 3 wt%) is found in Ultra Nano-Crystalline Diamond films (UNCD) (Zhang et al., 1999). These films are characterized by extremely small diamond grain sizes (5 to 50 nm) and grain boundaries of different configurations with widths of several Ångstroms. However, in these films the N concentration in *diamond grains* is much lower and does not exceed 0.2 wt% (Birrell et al., 2003). The excess nitrogen is present as $C_xN_y$ compounds (e.g., Zhang et al., 1999), probably at grain boundaries (Birrell et al., 2003; Zapol et al., 2001).

At the same time, detonation nanodiamonds (often called Ultra-Dispersed Diamonds – UDD) routinely produced by detonation of high-energy explosives in closed volumes (e.g., Dolmatov et al., 2007) contain considerable amounts of nitrogen (up to 3%) as well as H and O and are, in this respect, broadly similar to meteoritic nanodiamond. It is believed that UDD are formed behind the shock wave front on a time scale of 0.1 to 0.5 microseconds from carbon liberated from organic compounds (Staver et al., 1984). The average size of diamond grains varies from 3 to 8 nm, though a tail up to ~12 nm is often observed (Anisichkin et al., 1988; Vlasov et al., 2010). Theoretical calculations suggest that incorporation of nitrogen into a nanodiamond grain with a high surface/bulk ratio is energetically unfavorable (Barnard and Sternberg, 2007). However, recent high resolution studies of UDD by Electron Energy Loss Spectroscopy (EELS) showed that N is indeed located within the diamond grains, but is usually associated with extended defects such as grain and crystallite boundaries (Kvit et al., 2003, Turner et al., 2009, Vlasov et al., 2010). The results of an X-ray Photoelectron Spectroscopy (XPS) study of UDD are also interpreted as an evidence for the presence of N in diamond grains (Dementjev et al., 2007).

The results summarized above lead to the following conclusions. If a diamond is synthesized in conditions where crystal growth depends on the kinetics of the absorption-desorption of carbon atoms and on the N partition ratio with a surrounding medium, i.e. not very far from equilibrium, the nitrogen content of the *diamond* phase (i.e. strictly $sp^3$-carbon with Fd3m space group) will be low (< ~0.3–0.4 wt%). Such a situation is typical for terrestrial diamonds where the average N concentration is ~0.04 wt%, and the maximum N content is 0.3–0.35 wt% (Cartigny, 2005, Sellschop, 1992), as well for diamonds grown under High-Pressure-High-Temperature conditions (HPHT), which usually contain smaller amounts of N (~0.02 wt%). The highest reported values for HPHT diamonds are 0.33 wt% (Borzdov et al., 2002), i.e. similar to the maximum for natural macrodiamonds.

We note that the efficiency of nitrogen incorporation in the diamond lattice strongly depends on N speciation in the growth medium and on its partition ratio. Extensive studies of HPHT or CVD diamonds show that almost all nitrogen is incorporated as single substitutional atoms (Evans et al., 1992) and only a minor fraction (~$10^{-3}$) may be trapped as N-V-N complexes (Iakoubovskii et al., 2000a). Therefore, to achieve efficient incorporation, nitrogen in the growth medium should be present as single atoms or in weakly bound compounds.

When the diamond growth occurs in strongly non-equilibrium conditions, such as (but not limited to!) detonation synthesis, significant numbers of N atoms may be trapped inside the diamond grains, provided that N is abundant. Very fast linear growth rates in such cases makes the N incorporation rate largely insensitive to N speciation. Theoretical calculations of nitrogen behavior in detonation nanodiamonds (Barnard and Sternberg, 2007) and in UNCD (Zapol et al, 2001) show that it is much more favourable energetically to trap significant nitrogen concentrations at crystallite boundaries and extended defects than in the diamond lattice. This model appears to be supported also by experimental studies (Zhang et al., 1999, Vlasov et al., 2010). Metamorphic microdiamonds (sizes <300 microns) with high N content in the diamond lattice (up to 9600 ppm (Cartigny et al. 2004)) possibly deviate from this model but since their growth conditions and medium are not very well constrained, it is difficult to discuss them.

According to the proposed model, the MND grains enriched in N were probably formed in a non-equilibrium process in a medium with high nitrogen content. Verification of this hypothesis requires investigation of impurity-related defects in MND with special emphasis on nitrogen-related defects.

Up to now, no reliable evidence of nitrogen-related defects in MND has been reported using optics-based spectroscopy (luminescence, absorption). In the case of macroscopic diamonds, infra-red (IR) spectroscopy permits quantitative investigation of nitrogen in the diamond lattice. Based only on their positions, some peaks in IR spectra of meteoritic nanodiamonds were ascribed to nitrogen point defects (Hill et al., 1997; Braatz et al., 2000). Such assignment is usually unacceptable in a single-crystal field, where the assignment is verified by matching the whole defect-induced one-phonon IR absorption spectrum rather than a single peak.

One of the main difficulties in many spectroscopic studies of nanodiamonds is the ubiquitous presence of non-diamond carbon and functionalisation of the nanodiamond surfaces by various C, O, H groups such as carboxyl, carbonyl etc. (Lewis et al., 1989; Bernatowicz et al., 1990; Jiang and Xu, 1995). Especially in studies where absorption spectra are acquired, the signals due to the non-diamond phases are often much stronger than those of nanodiamond or its bulk defects and impurities. Importantly, the IR spectra of dispersed nanodiamonds never show absorption by the diamond lattice itself in the two-phonon region. Possibly, the lattice absorption is not observed due to its relative weakness (13 $cm^{-1}$) and the small amount of available meteoritic material (Hill et al., 1997; Braatz et al., 2000). Independently of the reason for the absence of the lattice absorption, the speculations about defect-related absorption in the one-phonon region are not justified. Moreover, as we will show below, nitrogen-related defects in MND are very different from those encountered in macrodiamonds and their IR absorption spectra are probably very different.

Here we report on the results of spectroscopic study of impurity-related defects in MND from Efremovka (CV3) and Orgueil (CI) chondrites using photoluminescence, Raman and X-ray absorption spectroscopy. These results are compared with literature data for synthetic nanodiamonds produced by various methods. Implications for the MND formation process are discussed.

## 2. SAMPLES AND METHODS

The nanodiamond samples were extracted from the Efremovka, CV3 and the Orgueil, CI chondrites (DE1, DE2 and OD7, OD13, respectively) using a well-established process involving multistage chemical treatment by HF, HF+HCl, KOH, $H_2O_2$, $K_2Cr_2O_7$, and $HClO_4$ at various temperatures and isolation of the colloidal nanodiamond at high pH values of the solution (Tang et al., 1988). The bulk samples DE1 and DE2 were obtained from aliquots of the same HF-HCl-resistant residue. However, for sample DE1 the aliquot was depleted by grains with density below 2 g/cm$^3$ as a result of heavy liquid sedimentation. For the bulk OD7 sample the acid-resistant residue was treated also by a mixture $H_3PO_4$+$H_2SO_4$ (1:1) at 220°C. Sample OD13 represents the fine-grained fraction remaining in the colloidal solution of the bulk OD7 sample aliquot after its ultracentrifugation (13500 g, 50 hours).

The difference between the grain sizes of the OD7 and OD13 samples appears to be correlated to differences in isotopic compositions of the carbon and xenon. According to the pyrolysis and combustion data of A. Verchovsky (Open University, Milton Keynes, UK) the $\delta^{13}C$ and $^{136}Xe/^{132}Xe$ values for the OD7 are -34.5±0.5‰ and 0.442±0.004, respectively. For OD13, they are -28.1±0.5‰ and 0.512±0.004. All the samples possess a translucent pale yellow color, typical for meteoritic colloidal diamond. The masses of the samples studied were in the 1 to 3 mg range.

It is important to note that the samples studied were indeed nanodiamond material which was unambiguously proved by X-ray diffraction, Raman and X-ray absorption spectroscopies (see below). In addition, the isotopic composition of nitrogen and noble gases is in the range reported previously for well-characterized samples of meteoritic nanodiamond (Russell et al., 1996, Huss and Lewis, 1994).

The average values of $\delta^{15}$N and C/N ratio for the DE1 and DE2 samples have been reported by Russell et al. (1996) are: 213±70 (plateau) and -229.3±8.6‰ (bulk), respectively. For the OD7 and OD13 samples these values, as measured by A. Verchovsky, are 75±10, -302±5‰ and 58±5, -300±5‰, correspondingly. The marked differences in the nitrogen concentrations and isotopic compositions between the Efremovka and Orgueil nanodiamonds are mostly due to different degrees of thermal metamorphism of their parent bodies (Russell et al., 1996).

A UDD sample treated by $HClO_4$ (4h at 220 $^{o}$C) was used for comparative purposes in our spectroscopic study. The nitrogen environment in a similar sample was studied by Vlasov et al. (2010) by Electron Electron Loss Spectroscopy (EELS).

The Efremovka MND samples were investigated using X-ray diffraction (XRD), Small-Angle X-ray scattering (SAXS), photoluminescence (PL) spectroscopy at room and at liquid He (LHe) temperatures and by Raman spectroscopy using visible (514 nm) and UV (244 nm) lasers at room temperature. The Orgueil samples were studied by luminescence, SAXS and imaging Near Edge X-ray Absorption Spectroscopy (NEXAFS).

To exclude sample heating during the Raman and PL measurements, the samples were embedded into pure copper, thus dramatically minimizing artifacts due to thermal effects. Our extensive experiments on nanodiamonds of various origins have shown the reliability of such sample preparation. The Raman spectra of the Efremovka nanodiamonds were obtained using a UV laser (244 nm) at low laser power (1 to 2.5 mW). The majority of the PL measurements were performed using a LABRAM HR spectrometer with an Ar+ laser (488 nm). The laser power was limited to 1 mW and the beam was focused to 2 to 5 microns on the surface of the MND samples. The accuracy of the determination of the positions of the peaks in the PL spectra was ±0.1 nm.

The nitrogen speciation in the MND from Orgueil and in the UDD was assessed by imaging Near Edge X-ray Absorption Fine Structure spectroscopy (NEXAFS). The measurements were performed at the WERA beamline at the ANKA synchrotron source, Karlsruhe. The samples of meteoritic and detonation nanodiamonds were prepared by drying drops of the colloidal solution of the MND and the water suspension of the UDD on a silicon or copper substrate. The C and N concentrations of the substrates were inferred from inert gas fusion and were far lower than the detection limit of the equipment employed. The samples were illuminated with monochromatic X-rays at an incident angle of 25° to their surface. Carbon and nitrogen absorption K-edges were analyzed. For the C K-edge the energy scans were performed between 280 eV and 315 eV, and for the N K-edge - between 395 eV and 430 eV. Up to 15 consecutive scans were performed at the nitrogen edge to improve the signal to noise ratio. The energy resolution at the C K-edge was 0.1 eV; at the N K-edge - 0.22 eV. The lateral distribution of the emitted electron intensity was mapped by means of a photoemission electron microscope (PEEM). The FOCUS IS-PEEM instrument used achieves a lateral resolution of ~100 nm at the energies employed.

The NEXAFS information on the chemical bonding and local electronic structure around the absorbing atom is obtained from the spectral X-ray scan. For each step of the X-ray energy, a PEEM image was recorded representing the laterally resolved electron emission $I(x, y)$ for a well defined excitation energy $E$. In this way, a four-dimensional data stack $\{I(x, y), E\}$ was produced, from which NEXAFS spectra from arbitrary sample areas were extracted by plotting the integrated image intensity $I$ of this specific area versus the photon energy $E$ (Berg et al., 2006). Using the imaging mode the areas with small agglomerates of nanodiamonds where no charging was present were selected and analysed. Standard corrections for dark and flat field were applied. The contribution from amorphous carbon on the X-ray optics was corrected by subtracting the spectra recorded at diamond-free parts of the silicon substrate.

Modeling of the NEXAFS spectra for several possible configurations of nitrogen in cubic and hexagonal diamond lattices was performed using feff8.2 software (Ankudinov et al., 1998). An atomic cluster with radius 6 Å around the central nitrogen atom was employed. The calculations were performed both for the unrelaxed lattice as well as for various degrees of lattice relaxation accounted for by preselected elongation of one or more C-N bonds. In addition, several calculations were performed for clusters containing a vacancy in the vicinity of the nitrogen impurity.

## 3. RESULTS

### 3.1. X-ray diffraction

Diamond reflections in the XRD patterns for MND are broad, which is due to small grain size and strain/defects (Fig. 1). The average size of the diamond crystallites (in the 111 direction) estimated from the width of the diffraction peaks is approximately 3 nm which is similar to the value obtained using transmission electron microscopy (2.6–2.8 nm, Lewis et al., 1989; Daulton et al., 1996). Similar values were also obtained by SAXS technique. In addition to obvious diamond reflections, a very weak peak corresponding to main graphite interlayer spacing is observed at 3.34 Å. This value is just slightly smaller than that for turbostratic graphite. From the relative intensity of the diamond 111 and graphite 002 peaks and corresponding values of efficiency of the reflections, the amount of graphite is estimated to be ~3–5 wt%. The presence of a broad halo at small diffraction angles (corresponds to high interplanar spacings) indicates that even the severe chemical treatment does not lead to complete elimination of amorphous carbon. However, it is impossible to rigorously estimate the relative fraction of the amorphous phase.

## 3.2. Spectroscopy

### *3.2.1. Raman spectroscopy*

Detection of the diamond Raman line in substances with a significant fraction of $sp^2$-bonded carbons is often challenging, since Raman scattering excited by visible radiation is more sensitive to π-bonding, which exists in carbon structures formed by $sp^2$-hybridised orbitals (C=C double bonds), whereas Raman scattering excited by UV radiation is more sensitive to σ-bonding. The σ-bonds are present in all carbon materials, but they dominate in carbons with $sp^3$-hybridization (C-C single bonds). The photon energy of UV radiation with a wavelength of 244 nm is equal to 5.08 eV, which is close to the energy band gap between electronic states of σ-bonds (5.47 eV) in diamond. Thus employment of the 244 nm laser creates near-resonance conditions enhancing the Raman cross-section of diamond carbon (Ferrari and Robertson, 2004; Mikhaylyk et al., 2005). In addition, the characteristic luminescence of nanodiamonds is very weak at UV excitation.

The UV Raman spectra of the DE1 and DE2 samples are shown in Figure 2. A broad, asymmetric, downshifted diamond peak is clearly observed along with broad bands due to graphite-like carbon and C=O bonds. The presence of the C=O bonds is due to oxygen, chemisorbed on the surfaces of nanodiamond grains (Lewis et al., 1989). The diamond peak is observed at 1327 $cm^{-1}$ in the spectrum of sample DE1; the spectrum of sample DE2 is noisier, but the diamond peak is located at approximately the same wavenumber: 1326 $cm^{-1}$. Spectra like those for DE1 and DE2 samples are typical for UDD (Vul’, 2006).

The first order Raman spectrum of diamond consists of a single line at 1332 $cm^{-1}$. At least two different phenomena which may act simultaneously could be responsible for the downshift of the Raman peak for the MND: (1) phonon confinement due to a decrease of diamond crystallites down to approx. 3 nm (Lipp et al., 1997), and (2) the presence of hexagonal diamond polytypes (hex-Dia). The average size of the grains in MND does not exceed 3 nm (Lewis et al., 1989, Daulton et al., 1996; our X-ray and SAXS measurements) and it is thus not surprising that size effects become pronounced in their Raman spectra. Though according to TEM, the fraction of purely hex-Dia particles is not very high, MND grains often contain stacking faults and other similar imperfections (Daulton et al., 1996). From a crystallographic point of view, many types of stacking faults in the cubic diamond lattice may be regarded as hexagonal polytypes. The Raman peaks of 2H and 6H diamond polytypes are located between 1319 and 1327 $cm^{-1}$ (Knight and White, 1989) Therefore, at least partially, the observed downshift may be due to such extended defects.

### *3.2.2. Photoluminescence*

The photoluminescence spectra of the samples studied at room- and liquid helium temperatures are broadly similar (Fig. 3a, see also Shiryaev et al., 2009). Such similarity indicates inhomogeneous, i.e. strain-related (e.g., Davies, 1970), line broadening as is expected for nanoparticles. The spectra are dominated by a broad band with a maximum around 560 nm (red emission) with shoulders around 530 and 610 nm. Such spectra are typical for UDD. The broad band probably represents overlapping signals from non-diamond carbon and from defects on the surfaces of diamond grains (Iakoubovskii et al., 2000b; Smith et al., 2008). It is not yet known if nanodiamonds synthesized by homogeneous nucleation and growth in the vapor phase (i.e., CVD-like processes) exhibit similar spectral features.

The PL spectra were recorded over a broad wavelength range, which allowed observation of a remarkable feature at 737 nm (1.681 eV) with a weak shoulder around 757 nm (Fig. 3b). Several luminescent centers in diamond have peaks in this wavelength range: the "silicon" center (see below), the neutral vacancy (the GR1 center), and a zero-phonon line from a radiation-related defect. However, the GR1 defect peaks at 741 nm, which is clearly different from the observed peak at 737 nm (the uncertainty of our measurements is ±0.1 nm). The radiation-related defect at 736 nm accompanies the GR1 defect, being much weaker. Therefore, the absence of the GR1 shows that the observed peak at 737 nm cannot be due to the radiation-related defect.

Presence of the main line (737 nm) and a prominent secondary line (757 nm) is an unambiguous manifestation of the well-known "silicon" defect, most probably consisting of a silicon ion in the divacancy (commonly called a silicon-vacancy complex or Si-V) (for a review see Zaitsev, 2001). The observed band is broad and the fine structure typical for this defect is not resolved due to size/strain effects and related inhomogeneous broadening. For comparison, a spectrum of a Si-V defect in nanocrystalline CVD film (Vlasov et al., 2009) is shown in Figure 3b. However, the observed band is still sufficiently narrow to be reliably ascribed to a defect in a crystalline lattice. Bands due to defects in (semi)amorphous local environment are invariably much broader. This defect is observed in Si implanted diamonds after annealing (Zaitsev et al., 1981), in CVD and UNCD films grown on Si substrate (Vavilov et al., 1980, Vlasov et al., 2009, Basov et al., 2009), and in synthetic (Clark et al., 1995) and natural macrodiamonds (Iakoubovskii et al., 2001). However, it is not observed in UDD, probably, due to the general absence of Si compounds in explosion chambers.

The relative intensity of the Si defect in spectra of the Efremovka nanodiamond is higher than in the Orgueil nanodiamonds (Fig. 3b). This may reflect different concentration of the Si impurity in these samples. A comparison of the relative intensities of the Si defects in Orgueil nanodiamond samples with slightly different median sizes (Fig. 3b) shows that in the sample made of slightly finer grains (OD13 vs OD7), the Si defect is less pronounced.

Unfortunately, the concentration of silicon in the MND can not be determined from the luminescence measurements. Initial attempts to obtain qualitative information from absorption spectroscopy were unsuccessful due to the very small amount of material available. The study of the MND dispersed on a substrate show that the Si-V defects are present in many grains including the smallest ones (1 to 2 nm in size). A very rough estimate of the possible concentration of silicon in MND suggests values around tens of ppm.

To the best of our knowledge, this is the first observation of the silicon defect not only in meteoritic, but in any type of dispersed nanodiamond.

Note that a defect present in all types of irradiated diamonds, the GR1, is absent in spectra of the samples studied here (see above). As follows from the name of this defect (G*eneral* R*adiation*), this defect appears after any type of irradiation. Its absence might imply that the irradiation of the MND was very moderate. However, this statement should be treated with caution, since the efficiency of GR1 generation strongly depends on the concentration and type of other defects present in the diamond and has been insufficiently studied for nanoparticles. Alternatively, the radiation defects may have been annealed before the formation of the parent bodies and prior to capture of the P3 noble gas component by the MND.

Spectra of MND samples kept in air for several years prior to the room temperature PL study sometimes showed existence of a broad PL band at 500 nm, rapidly vanishing even under low laser power. This feature is most likely due to some carbonaceous contamination, which builds up during storage. This contamination could also be responsible for the increase in $\delta^{13}C$ for carbon released from MND at low oxidation temperatures (Russell et al., 1996). The carbonaceous contamination is also observed in C NEXAFS spectra (see below).

### 3.3. Imaging NEXAFS and Photoelectron Emission Microscopy (PEEM)

Figure 4 shows a PEEM image of the MND from Orgueil at an incident energy of 293 eV, i.e. above the C absorption edge. The nanodiamonds are clearly visible as white (emitting) regions. Several nanodiamond-rich regions of the image stacks were selected and their C and N NEXAFS spectra were analyzed. Nanodiamond-free areas were used to correct for C contamination of the beamline optics.

Figure 5a shows the carbon edge for the UDD and for the MND from Orgueil (sample OD7). The spectra are typical for macro- and nanodiamonds (Morar et al., 1985, Birrell et al., 2003, Berg et al., 2008). They show the typical features of diamond: an absorption edge representing the conduction band edge, a C 1s core exciton, a series of structures resulting from a band of σ* states, and the second absolute band gap. The relative intensities of the features around the absorption edge could be slightly different from an ideal diamond due to difficulties in subtraction of the background in this region. Note the presence of a small peak at 285 eV indicating a small amount of $sp^2$-carbon (graphitic and/or aromatic), also typical for nanodiamonds. Similar to the Murchison nanodiamond (Berg et al., 2008), the sample from Orgueil shows a weak feature at 287 eV, which can be ascribed to semi-amorphous carbon on surfaces of nanodiamond grains, as also revealed by Raman spectroscopy. However, this feature is absent in the spectra of the UDD (Fig. 5a). The reason for such a discrepancy is not yet clear, but it may reflect differences in surface chemistry of UDD and MND.

The nitrogen absorption spectra (Fig. 5b) are relatively noisy. This clearly results from low concentrations of this element. In contrast to the spectrum of the UDD, a weak peak at 400 eV for the MND is obvious. This feature is most likely due to weakly $\pi$-bound nitrogen in a carbonaceous environment (Jimenez et al., 2001) probably corresponding to the (semi)amorphous outer shell which is also manifested in C NEXAFS and Raman spectra. The surface-bound nitrogen may have been chemisorbed during MND residence in space or during the separation procedure (Fisenko and Semjonova, 2006). The main absorption edge is observed at about 405 eV in the spectra of both the MND and the UDD. This position is indicative of N in $sp^3$-bonding configurations (see below). It should be noted that the N NEXAFS spectra of UNCD (Birrell et al., 2003) are similar to those observed in this study with the exception of the peak at 400 eV.

## 4. DISCUSSION

Optics-related methods such as IR-Vis-UV absorption and photoluminescence (PL) as well as Electron Paramagnetic Resonance (EPR) are extremely sensitive to nitrogen defects in the diamond lattice. The N concentration in diamond determined by these methods correlates quantitatively with data obtained by nuclear probes and inert gas fusion (Davies, 1999, Sellschop, 1992). However, nitrogen in nanodiamond produced by detonation or by CVD is difficult to observe by the spectroscopic techniques mentioned above, and only recently reliable detection of N-V defects have been reported for nanodiamond particles of 5–10 nm in size (Bradac et al., 2010; Smith et al., 2008; Vlasov et al., 2010). The published data on nanodiamond spectroscopy set an upper limit to nitrogen-related defects in diamond lattice of several hundred ppm at most (Iakoubovskii et al., 2000b; Orlinskii et al., 2011). Comparison of the concentration of paramagnetic nitrogen in nanodiamonds measured by EPR with mass-spectrometric determinations shows that the paramagnetic N atoms are always just a small fraction of the total N content of the samples (e.g., Orlinskii et al., 2011). The absence of clearly identified N-related signals in EPR spectra of MND implies that nitrogen in MND is predominantly non-paramagnetic (Braatz et al., 2000). In the current work, spectroscopic manifestations of the nitrogen-related defects in MND are also not observed (NEXAFS results are *not* considered yet).

Reconciliation of weakness of the spectroscopic manifestations of abundant structural N with the mass-spectrometry results showing N content of meteoritic and detonation nanodiamond as high as 1–2 wt% (Russell et al., 1996; Dolmatov, 2007) is not trivial. Note that the PL manifestations of the Si-V and the EPR spectrum of the proton-vacancy defects (the H1-defect, Braatz et al., 2000) in MND are fairly similar to manifestations of these defects in macrodiamonds if the strain-related broadening is taken into account. At the same time, the concentrations of the nitrogen-related defects observed are low. Several physical mechanisms could be responsible for considerable differences between the detection of the defects in nano- vs. bulk diamond by spectroscopic techniques.

1) The presence of 1–2 wt% of N in the diamond lattice inevitably produces considerable strains due to differences in C and N ionic radii. Configuration of the energy levels in a highly-strained nanodiamond grain as well as small distance between the defect and grain surface may allow efficient non-radiative de-excitation channels. As a result, some defects may possess very weak, if any, PL responses. In contrast to the NV, the "silicon" defect has very weak electron-phonon interaction, and the changes in the diamond density-of-states due to decreasing size should have smaller effect on spectral manifestations of the Si-V defect.

2) The structure of the N-related defects in a nanodiamond differs considerably from that in the bulk diamond.

The explanations provided are not mutually contradictory. However, the results of our NEXAFS study and previously published data (Birrell et al., 2003; Vlasov et al., 2010) suggests that differences in atomic structure of the N environments are very important.

Using the feff8.2 code (Ankudinov et al., 1998) we have performed theoretical modeling of N NEXAFS spectra in cubic and hexagonal diamond polytypes. Figure 5c shows modeled spectra for nitrogen configurations usually encountered in macrodiamonds: substitutional single and paired nitrogen atoms with several values of lattice relaxation, as well for some other configurations of several N atoms and vacancies. The modeled spectra differ from the experimental spectra. The spectra of nitrogen atoms surrounded by various arrangements of vacancies are characterized by fewer broad spectral features, making them qualitatively more similar to the observations. Results of our modelling are qualitatively similar to earlier calculations of EELS spectra of N in diamonds (Brydson et al., 1998).

The XPS spectra for N in UNCD and UDD suggest the presence of two or, perhaps, three types of N environment (Zhang et al., 1999; Dementjev et al., 2007). Although there is no unique interpretation of the N XPS and NEXAFS spectra in the C-N systems is not yet possible, the features observed may originate from nitrogen occupying two and three-coordinated structures with at least one of the nearest neighbours being also a two- or three-coordinated carbon (Titantah and Lamoen, 2007). This assignment is consistent with a suggestion that N is largely confined to grain- and intercrystallite boundaries and other extended defects in UNCD (Zapol et al., 2001; Birrell et al., 2001) and UDD (Vlasov et al., 2010).

Our N NEXAFS spectra of the MND can be represented as a superposition of a "sharp" feature at 400 eV and a broad absorption starting at 405 eV. The 400 eV peak is probably due to π-bonded, presumably chemisorbed N (see above). The broad absorption shows close resemblance to NEXAFS spectra of detonation nanodiamonds (Fig. 5) and UNCD (Birrell et al., 2003) and to N EELS spectra of UDD (Pichot et al., 2010). This similarity suggests that the average local environment of the main fraction of nitrogen impurity in MND, UDD and UNCD is similar. Following the XPS results (Zhang et al., 1999; Dementjev et al., 2007; Titantah and Lamoen, 2007), we suggest that N in these nanodiamonds is most likely distributed among several main atomic configurations, thus blurring the absorption spectra. However, the set of these configurations is not completely random and/or sample-dependent, which is indicated by the consistent shape of the absorption spectra. Therefore, the atomic-level N environment in MND is markedly different from the bulk diamond lattice and, therefore, its manifestations in IR absorption and luminescence must be different. This may explain the failure of reliable observation of N-related defects in MND (Braatz et al., 2000, Hill et al., 1997) by IR spectroscopy. Possibly, nitrogen in MND is largely segregated to extended defects such as crystallite and twin boundaries. According to a TEM study by Daulton et al. (1996), the concentration of twins in MND is not very high. However, this does not contradict the hypothesis that nitrogen in MND is mostly concentrated at extended imperfections, since the relative fraction of N-rich nanodiamond grains is unknown.

As discussed in the introduction, incorporation of high amounts of nitrogen (>1 wt%) into the diamond lattice is very difficult. N-rich nanocrystalline diamond films containing ~0.2 wt% of N are grown from a medium with high N (>10%) content. Linear growth rates of diamond crystallites in these films are usually low: <0.1 μm/h (May and Mankelevich, 2007).

Presumably, such slow growth leads to formation of N-poor nanograins joined by N-rich grain boundaries. In contrast, the linear growth rate of N-rich UDD is very high and may reach $10^5$ to $10^6$ μm/h. Presumably, the nitrogen atoms are simply buried by the carbon atoms. Preferential localization of N at the grain boundaries and between crystallites results from diffusion processes while the grain is still at high temperatures and from formation of N-poor "perfect" grains.

Application of this model of the formation of N-rich nanodiamonds to MND implies that these grains grew very fast in strongly non-equilibrium conditions and that their growth medium was rich in nitrogen in a form readily available for incorporation into the diamond structure.

Several modes of formation of MND with high amounts of nitrogen may be proposed.

1) Shock-induced transformation of the C-N-rich organics, for example, after supernovae explosion (Saslaw and Gaustad, 1969; Blake et al., 1988; Fisenko et al., 2001). Problems with this scenario are that several conditions should be achieved in the region of MND formation: a high density of starting matter and a high rate of temperature decrease. However, a density and temperature increase during the passage of a shock wave through a C-N rich gaseous medium may serve as an efficient trigger for fast disequilibrium growth of nanodiamonds, by, for example, a CVD-like process (see below). A somewhat similar process was employed in laser- or discharge-induced diamond growth (e.g., Buerki and Leutwyler, 1991).
2) Formation of nanodiamond in UV-irradiated ices and/or organics has been proposed by Nuth and Allen (1992). However, the growth rates currently obtained in experiments are low (Kouchi et al., 2005). The possibility of fast diamond growth by this mechanism, e.g., in a very intense flash of high energy photons, has still to be checked experimentally. In any case, this mechanism does not seem very efficient since it requires simultaneous destruction of several chemical bonds in a solid substance and subsequent rearrangement into the diamond structure. The probability of this process increases with fluence, but, at high UV fluences the stability of nanodiamonds against graphitization decreases (Butenko et al., 2008).

3) The CVD-like process is a very attractive mechanism (Lewis et al., 1989; Daulton et al., 1996). However, to achieve a high N content in the diamond grains, it should occur in environments rich in accessible N (in atomic form or as weakly bonded compounds) and proceed with high growth rates. The possibility of nanodiamond growth by CVD-like processes with high linear growth rates (up to thousands μm/h) has been demonstrated (Buerki and Leutwyler, 1991; Frenklach et al., 1991; Palnichenko et al., 1999).

In our view, the fast CVD-like mechanism is the most plausible. The CVD-like mechanism of formation of interstellar nanodiamonds is supported by microscopic observations of Daulton et al. (1996).

The N-rich grains of MND may represent just one of the possible sub-populations of MND. The presence of several populations of nanodiamond grains with somewhat different properties can be inferred from variations of the C, N and noble gases isotopic compositions observed during stepped pyrolysis and oxidation of nanodiamonds from different petrological and chemical classes of meteorites (Huss and Lewis, 1994; Russell et al., 1996; Verchovsky et al., 1998).

## 5. CONCLUSIONS

Phase compositions and defects in nanodiamonds separated from Efremovka, CV3 and Orgueil, CI chondrites were studied using Raman and photoluminescence spectroscopies at different excitation wavelengths, X-ray scattering and diffraction and X-ray absorption spectroscopy.

We show that the phase compositions of the meteoritic and the synthetic (detonation) nanodiamonds are similar, suggesting that both types of nanodiamonds presumably consist of a diamond core surrounded by (semi)amorphous and graphite-like carbon.

The nitrogen atomic environment appears to be broadly similar for meteoritic and detonation nanodiamonds as well as for Ultra-Nano-Crystalline CVD films. Most of the nitrogen in MND appears to be confined to lattice imperfections, such as crystallite/twin boundaries and other extended defects; the concentration of nitrogen in the MND lattice is low. This leaves little chance for success in the search for spectral manifestations of common N-related defects in meteoritic nanodiamonds.

The high nitrogen content in at least some MND grains suggests that their growth rate was high. One of the plausible formation mechanisms of MND is the CVD-like process in regions rich in accessible N (in atomic form or as weakly bonded compounds), possibly triggered by a shock wave. Other processes such as nanodiamond formation in irradiated organics can not be excluded. The subpopulation of N-rich nanodiamond grains could have formed both prior to and/or during Solar System formation. It seems unlikely that this subpopulation is a carrier of isotopically anomalous HL noble gases, which are presolar. This follows from negative correlation between contents of the nitrogen and the HL component of noble gases in nanodiamonds from meteorites subjected to different degrees of thermal metamorphism.

For the first time the silicon-vacancy complex (the “silicon” defect) has been observed in meteoritic nanodiamond by photoluminescence spectroscopy. Moreover, this is the first observation of this defect in dispersed nanodiamonds of any origin. Study of Si isotopic composition in nanodiamonds may help to understand their origin. To avoid contributions from the SiC grains, the investigation of Si isotopes should be performed on nanodiamond grains isolated from metamorphosed meteorites such as, e.g., Allende, CV3 and Efremovka, CV3, or a chemical treatment destroying SiC should be employed.

*Acknowledgements:* Discussions with Drs K. Iakoubovskii and V. Krivobok on interpretation of luminescence results are highly appreciated. We are grateful to Dr. A. Verchovsky for provision of the bulk sample of the Orgueil meteorite and of the results of isotopic measurements of Orgueil nanodiamonds. Comments of Dr. T. Daulton on the early version of the manuscript as well as comments of anonymous reviewers are highly appreciated. We highly appreciate help of Dr. A.N. Krot and Dr. T. Lafford in improving of English. This work was supported by RFBR grant 08-05-00745a and Program RAS #21. ANKA (Angstroemquelle Karlsruhe) synchrotron radiation facility is acknowledged for the provision of beamtime (project MR-97).

# References

Anisichkin V. F., Mal'kov I. Yu., Titov V. M. (1988) Diamond synthesis at dynamic loading of organic compounds. *Dokl. Akad. Nauk SSSR* **303**, 625-627.

Ankudinov A. L., Ravel B., Rehr J.J., and Conradson S.D. (1998) Real Space Multiple Scattering Calculation of XANES, *Phys. Rev.* **B 58,** 7565.

Barnard A. S. and Sternberg M. (2007) Can we predict the location of impurities in diamond nanoparticles? *Diam. Relat. Mater.* **16**, 2078-2082.

Basov A. A., Rahn M., Pars M., Vlasov I. I., Sildos I., Bolshakov A. P., Golubev V. G., and Ralchenko V. G. (2009) Spatial localization of Si-vacancy photoluminescent centers in a thin CVD nanodiamond film. *Phys. Status Solidi* **A206**, 2009-2011.

Berg T., Maul J., Erdmann N., Bernhard P., Schuppler S., Nagel P., Sudek C., Ott U., and Schönhense G. (2006) Coupling of imaging NEXAFS with secondary ion mass spectrometry for the chemical and isotopic analysis of presolar cosmic grains. *Anal. Bioanal. Chem.* **386**, 119-124.

Berg T., Marosits E., Maul J., Nagel P., Ott U., Schertz F., Schuppler S., Sudek C., and Schönhense G. (2008) Quantum confinement observed in the x-ray absorption spectrum of size distributed meteoritic nanodiamonds. *J. Appl. Phys.*, **104**, 064303.

Bernatowicz T. J., Gibbons P. C., and Lewis R. S. (1990) Electron energy loss spectrometry of interstellar diamonds. *Astrophys. J.* **359,** 246-255.

Birrell J., Gerbi J. E., Auciello O., Gibson J. M., Gruen D. M. and Carlisle J. A. (2003) Bonding structure in nitrogen doped ultrananocrystalline diamond. *J. Appl. Phys.* **93**, 5606-5612.

Blake D., Freund F., Bunch T. (1988) A comparison of Allende diamond with diamond from detonation soot. *Lunar Planet. Sci. Conf.* **XIX,** 94-95 (abstr).

Borzdov Yu., Pal’yanov Yu., Kupriyanov I., Gusev V., Khokhryakov A., Sokol A., Efremov A. (2002) HPHT synthesis of diamond with high nitrogen content from an $Fe_3N$–C system. *Diam. Relat. Mater.* **11**, 1863–1870.

Braatz A., Ott U., Henning Th., Jager C. and Jeschke G. (2000) Infrared, ultraviolet, and electron paramagnetic resonance measurements on presolar diamonds: Implications for optical features and origin. *Meteoritics & Planet. Sci.* **35**, 75-84.

Bradac C., Gaebel T., Naidoo N., Sellars M. J., Twamley J., Brown L. J., Barnard A. S., Plakhotnik T., Zvyagin A. V. and Rabeau J. R. (2010) Observation and control of blinking nitrogen vacancy centres in discrete nanodiamonds. *Nature Nanotechnology* **5**, 345-349.

Brydson R., Brown L. M., Bruley J. (1998) Characterising the local nitrogen environment at platelets in type IaA/B diamond. *J. Microsc.* **189**, 137-144.

Buerki P. R., and Leutwyler S. (1991) Homogeneous Nucleation of Diamond Powder by $CO_2$ Laser Driven Gas-Phase Reactions. *J. Appl. Phys,.* **69**, 3739–3744.

Butenko Yu. V., Coxon P. R., Yeganeh M., Brieva A. C., Liddell K., Dhanak V. R., Šiller L. (2008) Stability of hydrogenated nanodiamonds under extreme ultraviolet irradiation. *Diam. Relat. Mater.* **17**, 962–966.

Cartigny P., Chinn I., Viljoen K. S., Robinson D. (2004) Early Proterozoic (> 1.8 Ga) ultrahigh pressure metamorphism: Evidence from Akluilâk microdiamonds (NW Canada). *Science* **304**, 853-855.

Cartigny P. (2005) Stable isotopes and the origin of diamond. *Elements* **1,** 79-84.

Clark C. D., Kanda H., Kiflawi I. and Sittas G. (1995) Silicon defect in diamond. *Phys. Rev.* **B51,** 16681-16688.

Daulton T. L., Eisenhour D., Bernatowicz T., Lewis R., and Buseck P. (1996) Genesis of presolar diamonds: Comparative high-resolution transmission electron microscopy study of meteoritic and terrestrial nanodiamonds. *Geochim. Cosmochim. Acta* **60**, 4853-4872.

Davies G. (1970) No phonon line shapes and crystal strain fields in diamonds. *J. Phys. C: Sol. State Phys.* **3**, 2474-2486.

Davies G. (1999) Current problems in diamond: towards a quantitative understanding. *Physica* **B273-274**, 15-23.

Dementjev A. Maslakov K., Kulakova I., Korolkov V., Dolmatov V. (2007) State of C-atoms on the modified nanodiamond surface. *Diam. Relat. Mater.* **16** 2083–2086.

Dolmatov V. Yu. (2007) Detonation-synthesis nanodiamonds: synthesis, structure, properties and applications. *Russ. Chem. Rev.* **76,** 339-360.

Evans T. (1992) Aggregation of nitrogen in diamond. In: *The properties of natural and synthetic diamond* (ed. J. E. Field). Associated Press, London. pp. 259-290.

Ferrari A. C. and Robertson J. (2004) Raman spectroscopy of amorphous, nanostructured, diamond–like carbon, and nanodiamond. *Phil. Trans. R. Soc. Lond. A* **362,** 2477-2512.

Fisenko A. V., Verchovsky A. B., Semjonova L. F., and Pillinger C. T. (2001) Heterogeneity of the interstellar diamond in the CV3 meteorite Efremovka. *Astron. Lett.* **27,** 608-612.

Fisenko A. V., Russell S. S., Ash R. D., Semjonova L. F., Verchovsky A. and Pillinger C. T. (1992) Isotopic composition of carbon and nitrogen in the diamonds from the unequilibrated ordinary chondrite Krymka LL3.0. *Lunar Planet. Sci. Conf.* **23**, 365-366 (abstr).

Fisenko A. V. and Semjonova L. F. (2006) Populations of nanodiamond grains in meteorites from data on isotopic composition and content of nitrogen. *Solar System Research* **40,** 485-499.

Frenklach M., Howard W., Huang D., Yuan J., Sear K. E., Koba R. (1991) Induced nucleation of diamond powder. *Appl .Phys. Lett.* **59,** 546-549.

Hill H. G. M., D'Hendencourt L. B., Perron C., Jones A. P. (1997) Infrared spectroscopy of interstellar nanodiamonds from the Orgueil meteorite. *Meteoritics Planet. Sci.* **32**, 713-718.

Huss G. R. and Lewis R. S. (1994) Noble gases in presolar diamonds I: Three distinct components and their implications for diamond origins. *Meteoritics* **29,** 791-810.

Iakoubovskii K., Adriaenssens G. J., Vohra Y.K. (2000a) Nitrogen incorporation in diamond films homoepitaxially grown by chemical vapour deposition. *J. Phys.: Condens. Matter* **12** L519-L524.

Iakoubovskii K., Baidakova M. V., Wouters B. H., Stesmans A., Adriaenssens G. J., Vul' A. Ya., Grobet P. J. (2000b) Structure and defects of detonation synthesis nanodiamond. *Diam. Relat. Mater.* **9**, 861-865.

Iakoubovskii K., Adriaenssens G. J., Dogadkin N. N., and Shiryaev A. A., (2001) Optical characterization of 3H, H3 and Si-V centers in diamond. *Diam. Relat. Mater.* **10**, 18-26.

Jiang T. and Xu K. (1995) FTIR study of ultradispersed diamond powder synthesised by explosive detonation. *Carbon* **33**, 1663-1671.

Jimenez I., Gago R., Albella J. M., Terminello L. J. (2001) X-ray absorption studies of bonding environments in graphitic carbon nitride. *Diam. Relat. Mater.* **10**, 1170-1174.

Jin S. and Moustakas, T. D. (1994) Effect of nitrogen on the growth of diamond films. *Appl. Phys. Lett.* **65**, 403-405.

Knight D. S. and White W. B. (1989) Characterization of diamond films by Raman spectroscopy. *J. Mater. Res.* **4**, 385-393.

Krauss A. R., Gruen D. M., Zhou D., McGauley T. G., Qin L. C., Corrigan T., Auciello O., Chang R. P. H. (1998) Morphology and electron emission properties of nanocrystalline CVD diamond thin films. *Mater. Res. Soc. Symp. Proc.* **495,** 299-310.

Kouchi A., Nakano H., Kimura Y., and Kaito C. (2005) Novel route for diamond formation in interstellar ices and meteoritic parent bodies. *Astrophys. J.* **626,** L129–L132.

Kvit A. V., Zhirnov V. V., Tyler T., Hren J. J. (2003) Aging effect and nitrogen distribution in diamond nanoparticles, *Composites* **B35,** 163–166.

Lewis R. S., Anders E., and Draine B. T. (1989) Properties, detectability and origin of interstellar diamonds in meteorites. *Nature* **339**, 117-121.

Lipp M. J., Baonza V. G., Evans W. J., and Lorenzana H. E. (1997) Nanocrystalline diamond: Effect of confinement, pressure, and heating on phonon modes. *Phys. Rev.* **B56,** 5978-5984.

May P. W. and Mankelevich Yu. A. (2007) The mechanism for Ultrananocrystalline Diamond Growth: Experimental and Theoretical Studies. *Mater. Res. Soc. Symp. Proc.* **956,** 0956-J07-04.

Morar J. F., Himpsel F. J., Hollinger G., Hughes G., and Jordan J. L. (1985) Observation of C 1s core exciton in diamond. *Phys. Rev. Lett.* **54**, 1960-1963.

Mykhaylyk O. O., Solonin Yu. M., Batchelder D. N., Brydson R. (2005) Transformation of nanodiamond into carbon onions: A comparative study by high-resolution transmission electron microscopy, electron energy-loss spectroscopy, x-ray diffraction, small-angle x-ray scattering, and ultraviolet Raman spectroscopy. *J. Appl. Phys.* **97**, 074302-1-15.

Newton J., Bischoff A., Arden R. D., Franchi I. A., Geiger T., Greshake A. and Pillinnger C. T. (1995) Acfer 094, a uniquely primitive carbonaceous chondrite from the Sahara. *Meteoritics* **30**, 47-56.

Nuth J. A. and Allenn J. E. Jr. (1992) Supernovae as sources of interstellar diamonds. *Astrophysics Space Sci.* **196,** 117-123.

Orlinskii S. B., Bogomolov R. S., Kiyamova A. M., Yavkin B. V., Mamin G. M., Turner S., van Tendeloo G., Shiryaev A. A., Vlasov I. I. and Shenderova O. (2011) Identification of Substitutional Nitrogen and Surface Paramagnetic Centers in Nanodiamond of Dynamic Synthesis by Electron Paramagnetic Resonance. Nanosci. & Nanotech. Lett. **3**, 63–67.

Pichot V., Stephan O., Comet M., Fousson E., Mory J., March K., and Spitzer D. (2010) High Nitrogen Doping of Detonation Nanodiamonds. *J. Phys. Chem. C.* **114**, 10082–10087.

Russell S. S., Arden J. W., and Pillinger C. T. (1996) A carbon and nitrogen isotopic study of diamond from primitive chondrites. *Meteoritics and Planet. Sci.* **31,** 343-355.

Samlenski R., Haug C., Brenn R., Wild C., Locher R., and Koidl P. (1995) Incorporation of nitrogen in chemical vapor deposition diamond. *Appl. Phys. Lett.* **67**, 2798-2800.

Saslaw W. C. and Gaustad J. E. (1969) Interstellar dust and diamond. *Nature* **221,** 160-162.

Sellschop J. P. F. (1992) Nuclear probes in the study of diamond. In: *The properties of natural and synthetic diamond* (ed. J. E. Field), Associated Press, London. pp. 81-180.

Shiryaev A. A., Fisenko A. V., Krivobok V. S., Vlasov I. I., Semjonova L. F. (2009) Nitrogen in Meteoritic Nanodiamonds: Lattice Impurity in Diamond Core or a Constituent of an Associated Carbonaceous Phases? *Lunar Planet. Sci. Conf.* **40,** 1317 (abstr).

Smith B. R., Inglis D. W., Sandnes B., Rabeau J. R., Zvyagin A. V., Gruber D., Noble C. J., Vogel R., Osawa E., Plakhotnik T. (2008) Five-nanometer diamond with luminescent nitrogen-vacancy defect centers. *Small* **5,** 1649-1653.

Staver A. M., Ershov A. P. and Lyamkin A. I. (1984) Study of detonations in condensed explosives by conduction methods. *Combustion, Explosion, and Shock Waves*, **20**, 320-324.

Tang M., Lewis R. S., Anders E., Grady M. M., Wright I. P., and Pillinger C. T. (1988) Isotopic anomalies of Ne, Xe and C in meteorites. I. Separation of carriers by density and chemical resistance. *Geochim. Cosmochim. Acta* **52**, 1221-1234.

Titantah J. T. and Lamoen D. (2007) Carbon and nitrogen 1s energy levels in amorphous carbon nitride systems: XPS interpretation using first-principles. *Diam. Relat. Mater.* **16,** 581–588.

Turner S., Lebedev O. I., Shenderova O., Vlasov I. I., Verbeeck J., and van Tendeloo G. (2009) Determination of Size, Morphology, and Nitrogen Impurity Location in Treated Detonation Nanodiamond by Transmission Electron Microscopy. *Adv. Funct. Mater.* **19**, 2116–2124.

Vavilov V. S., Gippius A. A., Zaitsev A. M., Deryagin, B. V., Spitsyn B. V. and Aleksenko A. E. (1980) Investigation of the cathodoluminescence of epitaxial diamond films. *Sov. Phys. Semicond.* **14**, 1078-1079.

Verchovsky A. B., Fisenko A. V., Semjonova L. F., Wright I. P., Lee M. R., and Pillinger C. T. (1998) Noble gas isotopes in grain size separates of presolar diamonds from Efremovka. *Science* **281**, 1165-1168.

Vlasov I. I., Barnard A. S., Ralchenko V. G., Lebedev O. I., Kanzyuba M.V., Konov V. I., and Goovaerts E., (2009) Nanodiamond photoemitters based on strong narrow-band luminescence from silicon-vacancy defects, *Adv. Mater.*, **21**, 808-812.

Vlasov I. I., Shenderova O., Turner S., Lebedev O. I., Basov A. A., Sildos I., Rähn M., Shiryaev A. A., and van Tendeloo G. (2010) Nitrogen and Luminescent Nitrogen-Vacancy Defects in Detonation Nanodiamond. *Small,* **6**, 687-694.

Vul' A. Ya. (2006) Characterisation and physical properties of UNCD particles. In *Ultrananocrystalline Diamond* (Eds: O. Shenderova, D. Gruen), William-Andrew, Norwich, NY. 379-404.

Zaitsev, A.M. (2001) *Optical Properties of Diamond*. Springer, Berlin.

Zaitsev A. M., Vavilov V. S. and Gippius A. A. (1981) *Sov. Phys .Lebedev Inst. Rep.* **10,** 15-20.

Zapol P., Sternberg M., Curtiss L. A., Frauenheim T., Gruen D. M. (2001) Tight-binding molecular-dynamics simulation of impurities in ultrananocrystalline diamond grain boundaries *Phys. Rev.* **B65**, 045403.

Zhang G. F., Geng D. S., Yang Z. J. (1999) High nitrogen amounts incorporated diamond films deposited by the addition of nitrogen in a hot-filament CVD system. *Surface Coatings Technology* **122,** 268–272.

Zhdankina O. Yu., Zadneprovskii B. I., Artemenko V. V., Kulakova I. I., Rudenko A. P. (1986) Influence of concentration of paramagnetic nitrogen on oxidation kinetics of synthetic diamonds of various morphologies. *Bull. Moscow State Univ. Chem.* **27**, 293-297.

## Figure captions

**Figure 1.** X-ray diffraction pattern of nanodiamond from Efremovka chondrite. Monochromatic molybdenum radiation was employed. Numbers show Miller indexes of the diamond reflections.

**Figure 2.** Raman spectra of meteoritic nanodiamond from Efremovka meteorites recorded using ultra-violet (244 nm) excitation. The Orgueil MND are marked as OD; the Efremovka MND are marked as DE; the synthetic detonation nanodiamond as UDD.

**Figure 3.** Photoluminescence spectra of meteoritic nanodiamonds recorded at 488 nm excitation at room temperature. Typical spectrum of the nanodiamond made by detonation is also shown. The arrow points to the main and secondary lines of the Si-V defect. All spectra are normalized and are displaced vertically for clarity. The figure B shows enlarged spectral region where the Si-V defect is observed. The curves are *not* displaced vertically, but are normalized to the maximum of the band at 550-600 nm. For comparison a spectrum of a Si-V defect in nanocrystalline CVD film is shown (Vlasov et al., 2009).

**Figure 4.** Photoemission (PEEM) image of a Si plate with precipitated nanodiamond from Orgueil meteorite (bright white bands) at the C absorption edge. The field of view is 37 microns.

**Figure 5.** NEXAFS spectra at carbon (A) and nitrogen (B) absorption edges of the meteoritic nanodiamond from Orgueil (MND) and detonation (UDD) nanodiamond. Figure C shows calculated NEXAFS spectra of various configurations of nitrogen in cubic (c-dia) and hexagonal (h-dia) diamond lattices. The curves are displaced vertically for clarity; horizontal axis shows energy relative to the absorption edge. Marked differences between the shapes of the experimental and modeled absorption spectra are obvious.

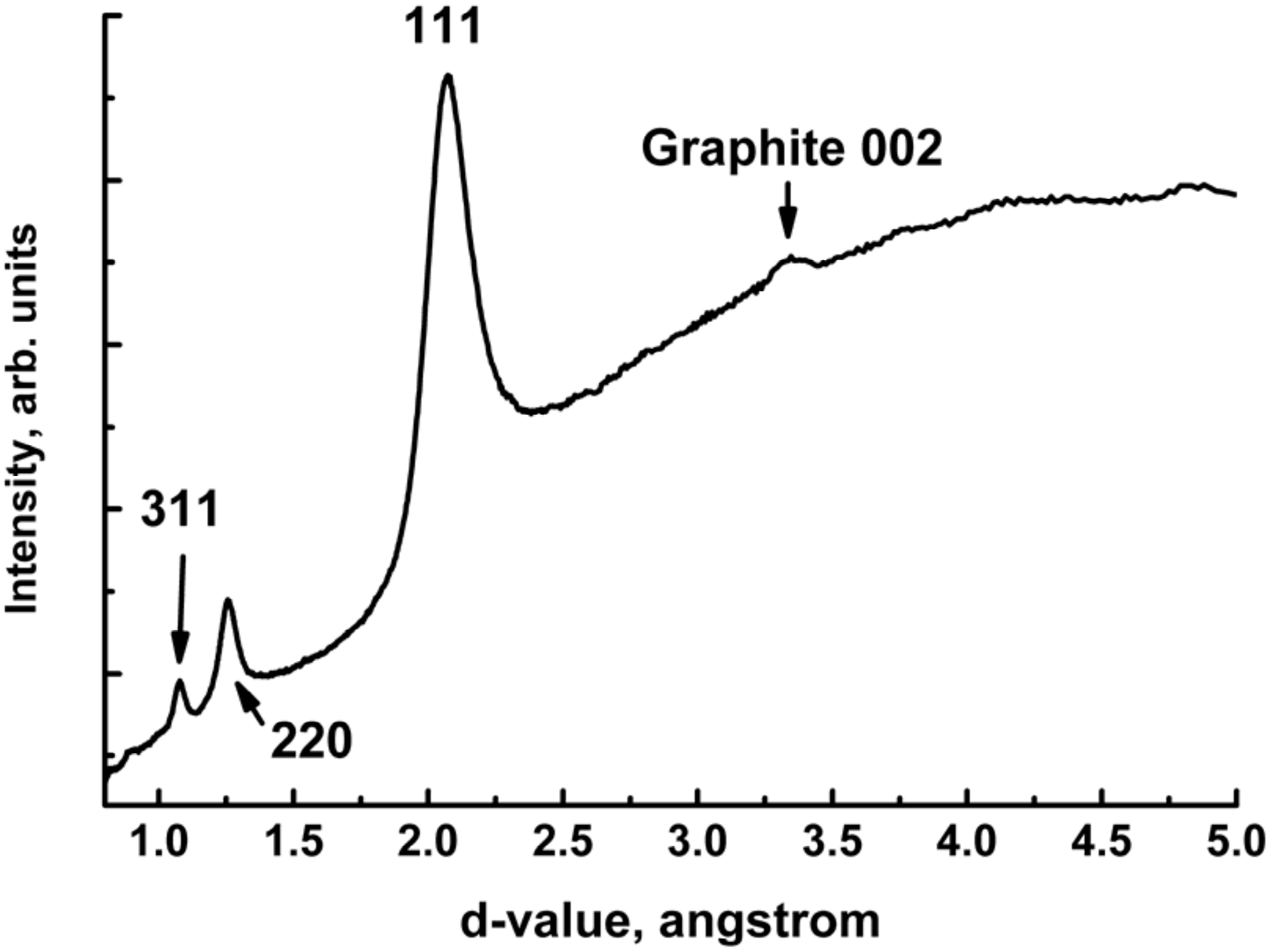

**Figure 1.**

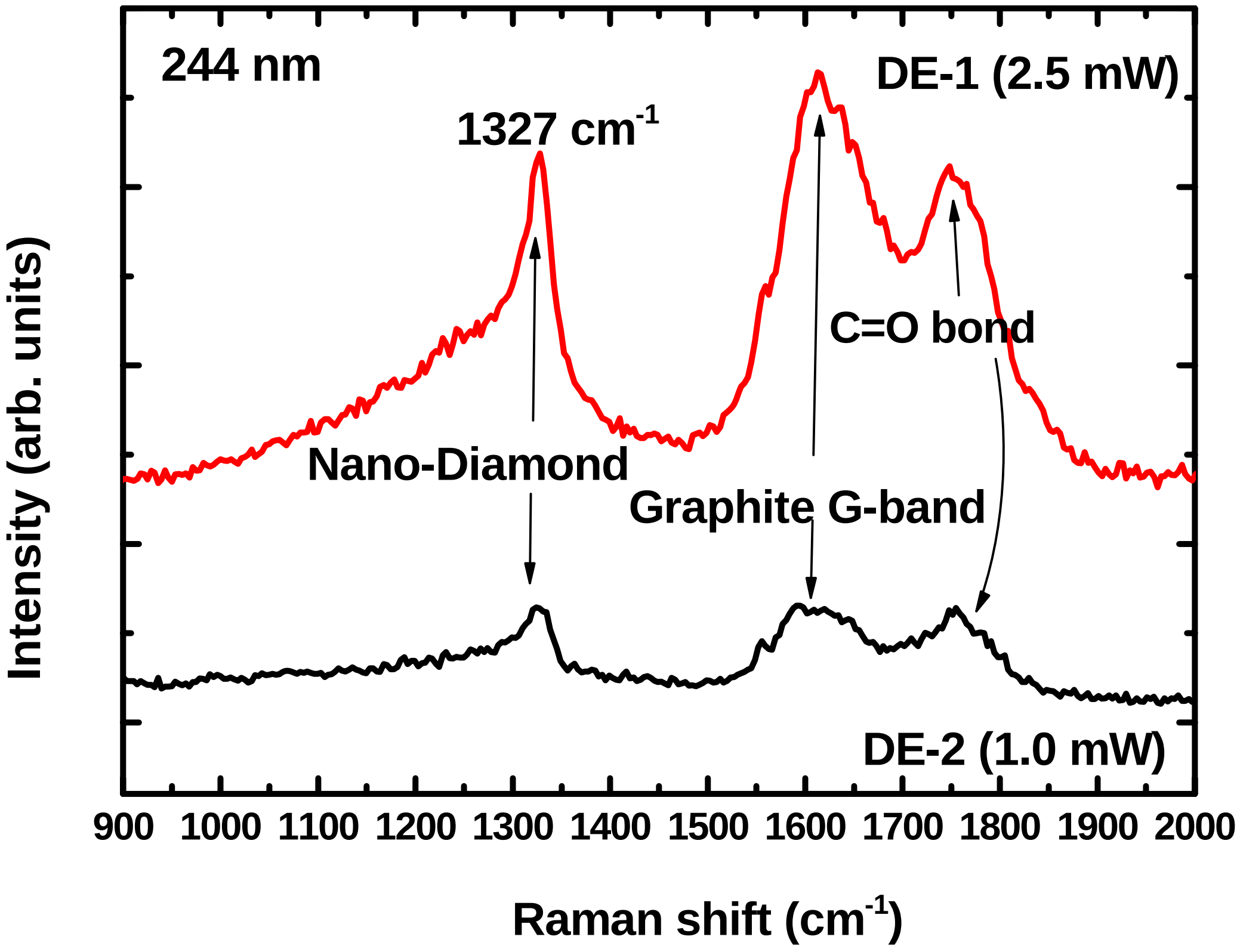
244 nm
DE-1 (2.5 mW)
1327 cm-1
C=O bond
Nano-Diamond
Graphite G-band
DE-2 (1.0 mW)
Intensity (arb. units)
900 1000 1100 1200 1300 1400 1500 1600 1700 1800 1900 2000
Raman shift (cm-1)

**Figure 2.**

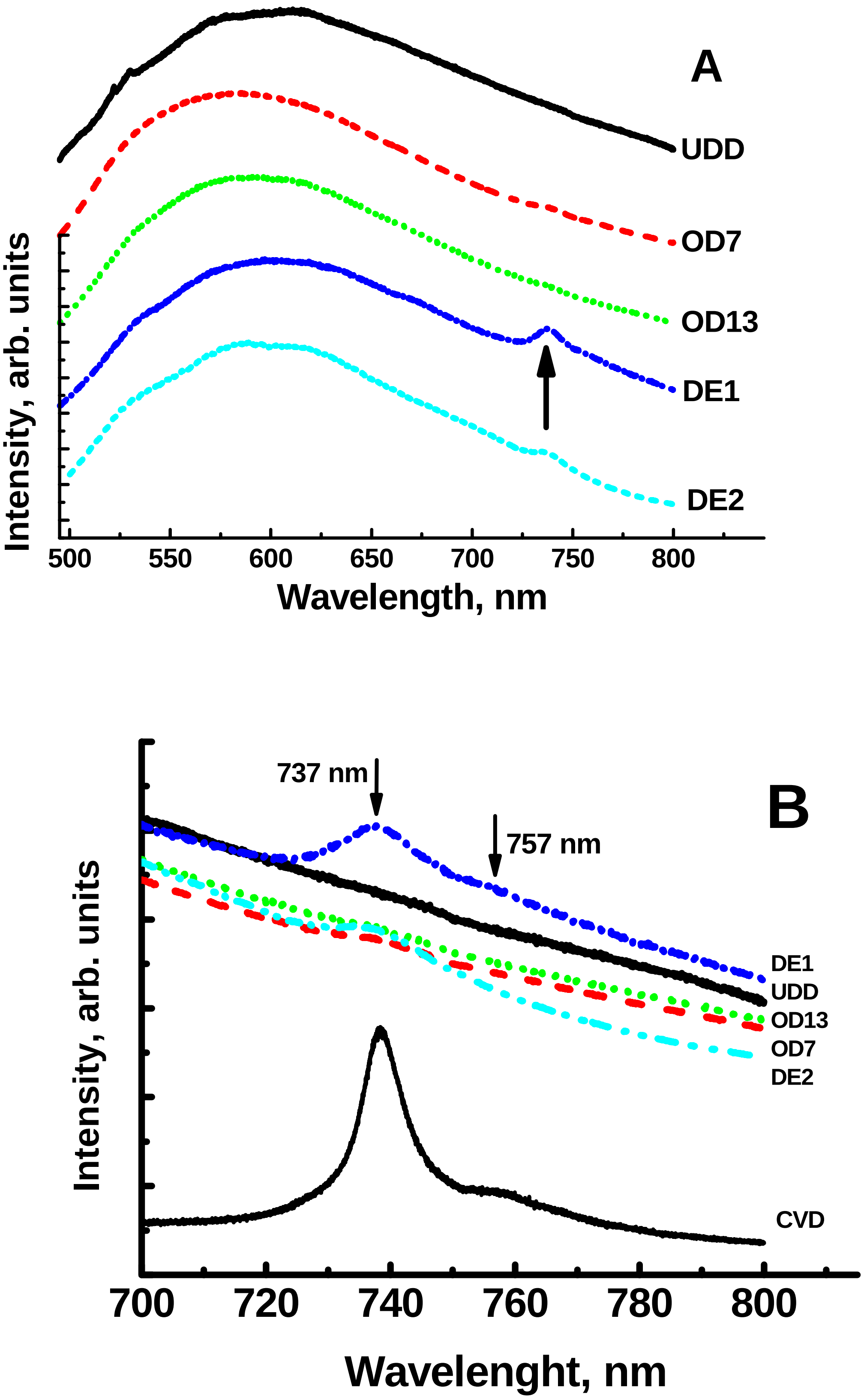

**Figure 3.**

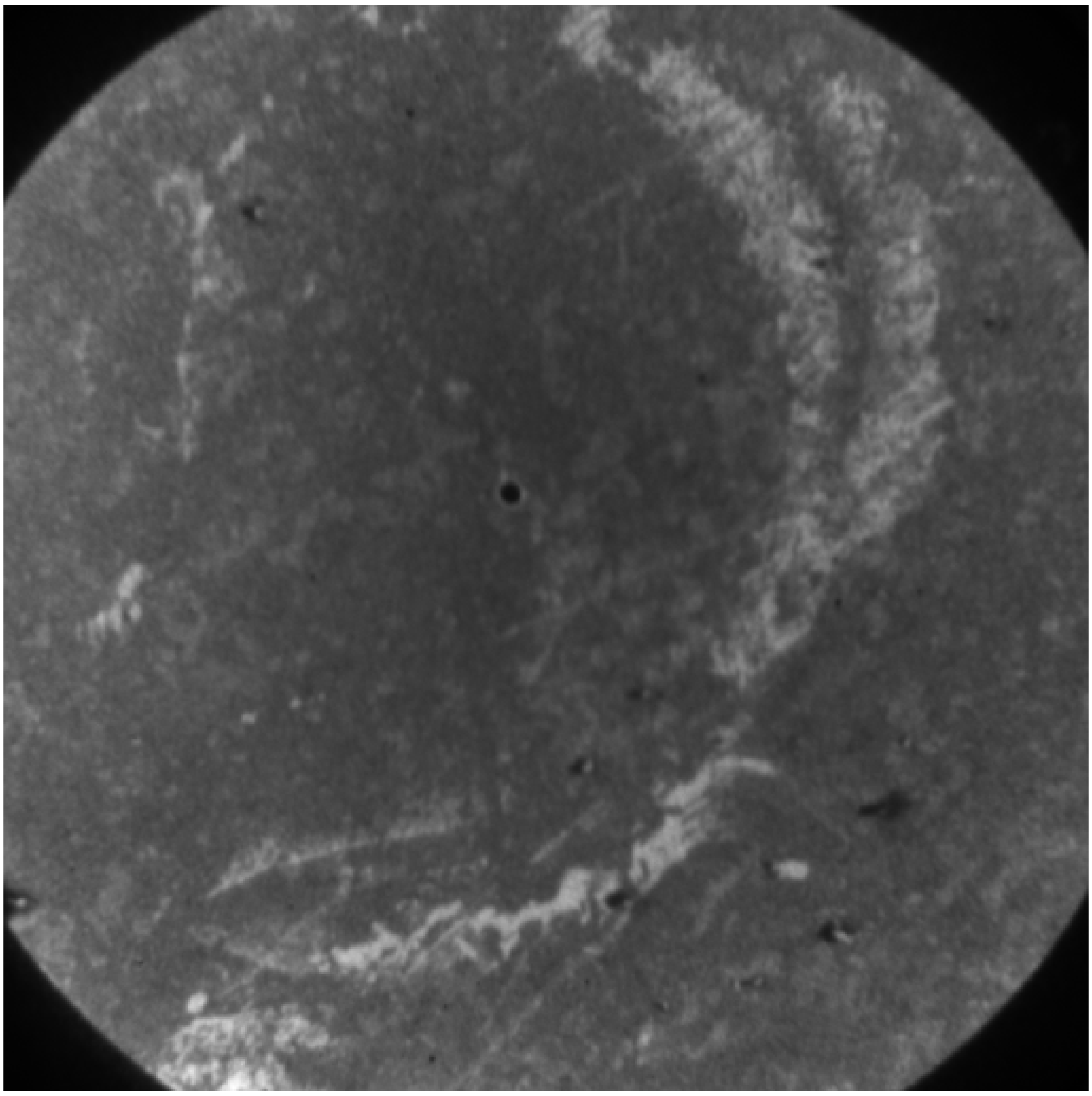

**Figure 4.**

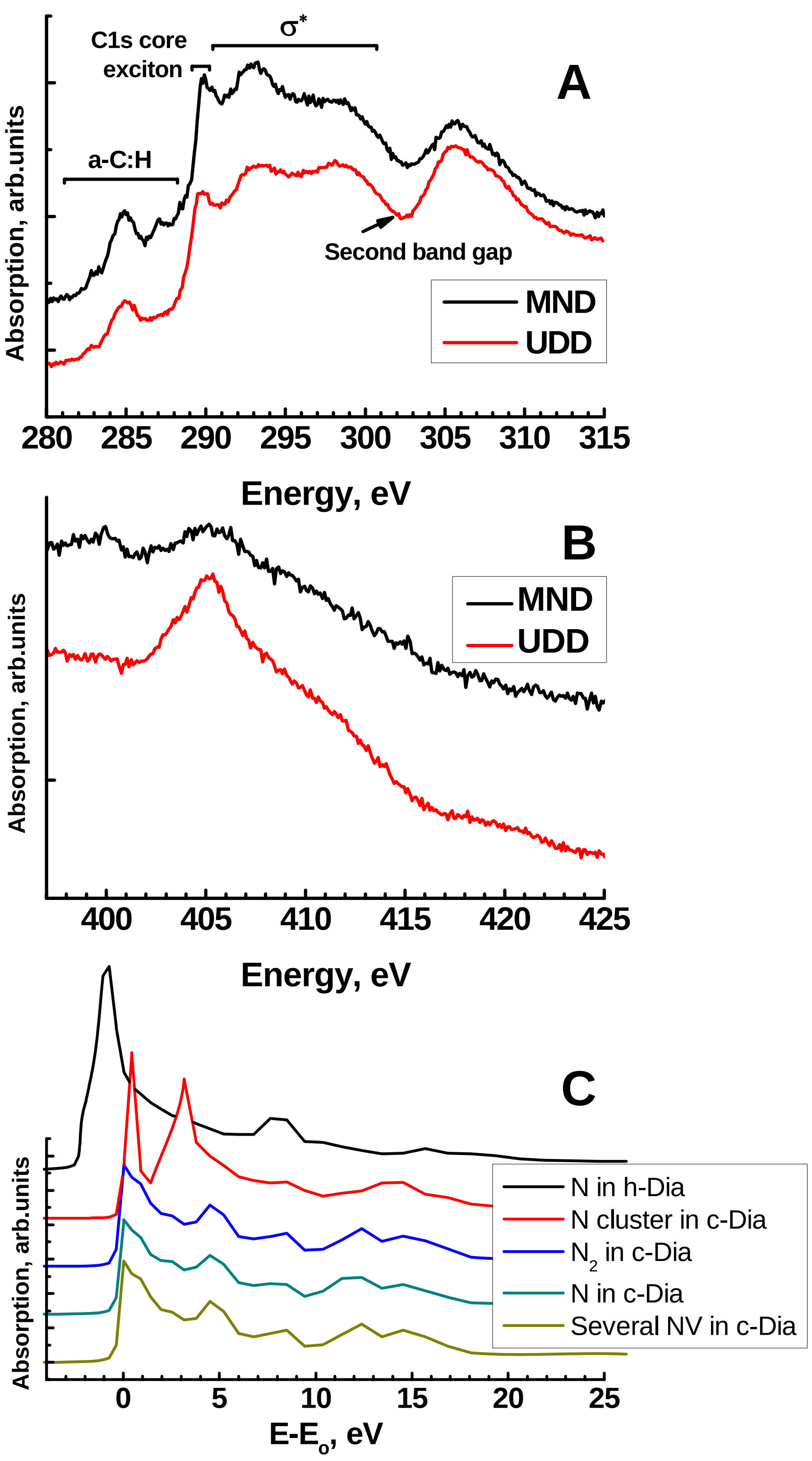

770
771 **Figure 5.**